\documentclass{aastex631}

\shorttitle{Gravitational Redshift in Open Clusters}
\shortauthors{Gutiérrez and Ramos-Chernenko}

\begin{document}

\title{Detection of Gravitational Redshift in Open Cluster non-degenerate stars}

\author{Carlos M. Gutiérrez}
\affiliation{Instituto de Astrofísica de Canarias, \\
Vía Láctea S/N, E-38205 La Laguna, Tenerife, Spain}

\author{Nataliya Ramos-Chernenko}
\affiliation{Instituto de Astrofísica de Canarias, \\
Vía Láctea S/N, E-38205 La Laguna, Tenerife, Spain}
\affiliation{Universidad de La Laguna, \\ 
 E-30206 La Laguna, Tenerife, Spain}

\begin{abstract}
A key observational prediction of Einstein's Equivalence Principle is that light undergoes redshift 
when it escapes from a gravitational field. Although astrophysics provides a wide variety 
of physical conditions in which this redshift should be significant, 
till very recently the observational evidence for this gravitational effect was limited to 
the light emitted by the Sun and white dwarfs. \textit{Gaia}-DR2 astrometric and kinematic 
data, in combination with other spectroscopic observations, provides a test bench to validate 
such predictions in statistical terms. The aim of this paper is to analyze several thousand 
main-sequence and giant stars in open clusters (OCs) to measure the gravitational redshift effect.
Observationally, a spectral shift will depend on the stellar mass-to-radius ratio as expected from the theoretical estimation of relativity. 
After the analysis, the obtained correlation coefficient between theoretical predictions and observations for 28 (51) OCs is $a= 0.977 \pm 0.218$ ($0.899 \pm 0.137$). 
The result has proven to be statistically robust and with little dependence on the details
 of the methodology or sample selection criteria. This study represents one of the more extensive validations of a fundamental prediction of gravity theories.

\end{abstract}


\section{Introduction} \label{sec:intro}


According to the predictions of the Einstein's Equivalence Principle (EEP) the light emitted in the stellar photospheres suffers a redshift when escaping its gravitational field, which is directly correlated to the mass-to-radius ratio of the observed star. In numbers, the measured effect is of the order of several
 hundred m\,s$^{-1}$ for main-sequence stars and tens of times smaller for giants. For white dwarfs the gravitational effect is 
several tens of km\,s$^{-1}$ owing to their comparatively small radius and extremely massive cores. That explains why the 
first attempts at detecting this relativistic effect was achieved in these types of stars 
by \citet{1967ApJ...149..283G} and \citet{1972ApJ...177..441T}.

The proximity of the Sun allows high-quality observations that are 
unattainable for other stars. However, since the Sun is a main-sequence star, the 
gravitational effect is relatively small and the observations need to consider the shift 
associated with the convective nature of the photosphere. In the solar spectrum, for example, 
the absorption line wavelength is blueshifted by about 400 m\,s$^{-1}$ after correction for 
the known gravitational redshift \citep{2003A&A...401.1185L}. In general, the precise amount 
of convective shift depends on the strength of the absorption line, i.e. stellar metallicity, 
surface gravity, and spectral type, or effective temperature \citep{2002ApJ...566L..93A, 2013A&A...550A.103A}. 
Several measurements have been made of the Sun \citep{2014MNRAS.443.1837R}, the most precise estimate being 
performed by \citet{2020A&A...643A.146G}, who take into account all these effects to
 determine the Sun's gravitational redshift at 633.1 m\,s$^{-1}$ with an uncertainty of $\sim $ 1\%.

Extending the test to other non-degenerate stars is of interest in order to test the EEP and the accuracy of stellar models. Nevertheless, as outlined above, such experiments 
are difficult: on the one hand, the hydrodynamic effect of the stellar photosphere produces a 
counterbalancing effect that tends to cancel out the relatively small gravitational redshift effect of 
the typical main-sequence star; on the other hand, obtaining data with high enough quality requires 
considerable effort in terms of observing time and instrumentation of high spectral resolution. 
These considerations explain the relatively low number of such attempts \citep{1997A&A...322..460N, 2019MNRAS.483.5026L, 2002A&A...381..446M}. Among these, it is worth mentioning the 
work by \citet{2011A&A...526A.127P}, who measured the shift of spectral lines for a sample of 
144 member stars in the M67 open cluster. The aim was to measure a radial velocity shift 
difference between the main-sequence and giant stars. The detection was unsuccessful, so the 
authors conclude that the convective shift probably mimicked the gravitational effect. 
More recently, \citet{2019ApJ...871..119D} used \textit{Gaia} radial velocity data to measure 
the redshift in the main-sequence stars and found values that are much smaller than the 
theoretical predictions. Again, the authors attributed this deficiency to the convective 
blueshift effect. A more successful attempt to detect the gravitational effect was carried out 
by \citet{2021arXiv210201079M} in non-degenerate stars. These authors measured the differences 
in velocities between the components of pairs of co-moving stars formed by dwarf--giant binaries. 
Instead of assuming the stellar hydrodynamical effect, they used a parametric model 
\cite{2013A&A...550A.103A} to estimate the convective shift. The final results provide 
compelling evidence for the compatibility between observations and theoretical predictions. Actually, measuring the gravitational redshift in galaxy clusters (e.g., \citealt{1995A&A...301....6C, 2011Natur.477..567W, 2013MNRAS.435.1278K}) and quasars \citep{Mediavilla_2021} could also be used to test General Relativity versus alternative theories of gravitation.

Nowadays, the availability of huge databases with astrometric, photometric and spectroscopic information, such as \href{https://www.sdss.org}{SDSS} and \textit{Gaia} \citep{2016A&A...595A...1G}, has provided sufficient objects to detect the gravitational redshift in a statistical way. 
By using a sample of member stars of open clusters catalogued by \textit{Gaia}, we aim to 
detect the effect of stellar gravitational redshift in non-degenerate stars. The 
data analyzed contain several thousand non-degenerate stars covering a wide range of mass and 
radius that potentially allows us to detect relative variations in their gravitational redshift. Unfortunately, the relatively weak gravitational fields existing in non-degenerated stars and the accuracy of the data used in this work does not allow to discriminate between different gravitational theories.

The paper is structured in a standard way. After this introduction, Section 2 describes 
the sample and methodology used; in Section 3 are presented the analysis, main results, and 
their interpretation. Finally, conclusions and planned future work are given in Section 4.


\section{Methodology and Sample} \label{sec:method}

For this study, we use a sample of stars in open clusters (OCs) of our galaxy. 
OCs are young stellar clusters which have been formed from the same molecular cloud, 
so their members share roughly the same age and metallicity \citep{2020SSRv..216...69A}. 
As co-moving systems bounded by mutual gravitational attraction, their
 the stellar peculiar velocity in the radial direction may be assumed to be random and 
to cancel out on average for each OC. 

\begin{figure}[tb!]
\plotone{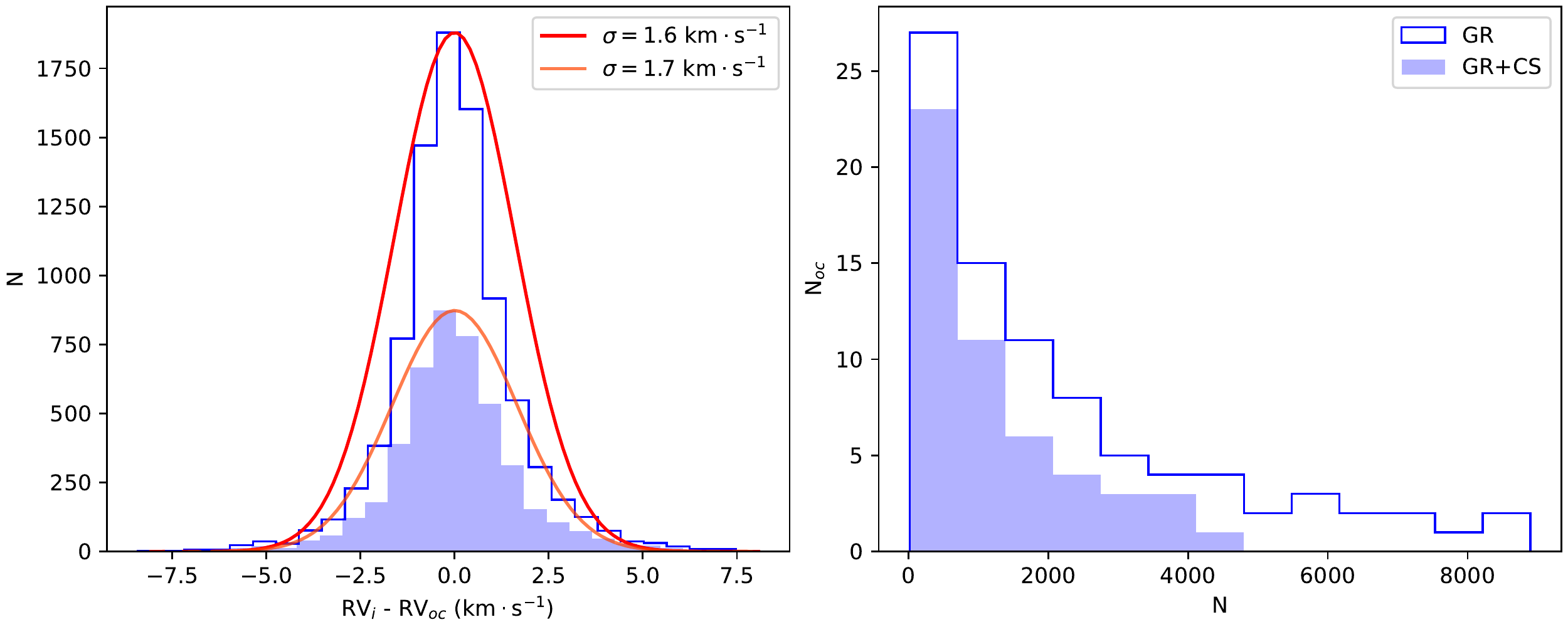}
\caption{\textit{Left panel}: Distribution of the stellar radial velocities 
($RV_i$) relative to its host cluster ($RV_{\rm OC}$). The solid lines are the 
Gaussian distribution fits for each sample, with corresponding standard deviations 
($\sigma$) depicted in the label. $N$ denotes the number of stars. \textit{Right panel}: 
Histogram distribution of the member stars ($N$) population in open cluster ($N_{\rm OC}$). 
In both panels, the step-type histogram corresponds to the full data set of 86 clusters 
and 8899 stars, labeled GR; the filled histogram corresponds to a subset of 51
 clusters and 4464 stars for which it is possible to estimate the convective shift, 
labeled GR+CS. See main text for details. \label{fig:histRV}}
\end{figure}

We took advantage of the largest radial velocity ($RV$) catalogue for stars in OCs by 
\citet{2021A&A...647A..19T}. It gathers nearly 30\,000 $RV$ measurements of 1382 OC 
members based on \textit{Gaia}-RVS and ground-based surveys (see references within). 
The complete OCs catalogue and the initial membership probability assessment came mostly 
from \citet{2020A&A...640A...1C} and \citet{2020A&A...633A..99C}, both based on \textit{Gaia} DR2 astrometry. 
The initial 1382 OCs data were completed with stellar parameters from the \textit{Gaia} 
DR2 based StarHorse catalogue \citep{2019A&A...628A..94A}. This provides estimations of 
photo-astrometric distances, extinctions, and other astrophysical parameters such as
 effective temperature ($T_{\rm eff}$), surface gravity ($\log g$) and metallicity ([Fe/H]).
 The final dataset contains 1342 OCs and 21\,403 stars classified as OC members with a
 minimum membership probability of 0.4, a threshold estimated by the authors. That 
constitutes the initial data set used in this paper.

In order to select an appropriate subsample for this work, only stars with uncertainties
up to 2 km\,s$^{-1}$ in $RV$ measurements were considered. This first restriction removes 
24\% of the initial sample objects from the analysis, and was motivated by the need to discard not 
only objects with poor measurements but also those that are part of binary systems, whose velocities could 
be dominated by the binary orbit itself. 

To constrain the stellar membership of the host cluster provided by \citet{2018A&A...619A.155S}, 
radial velocities were used as an additional criterion \citep{1997A&A...322..460N}. It was assumed that the radial 
velocities of the stars in a given OC correspond to a sample of a Gaussian distribution 
centred on $RV_{\rm OC}$ with a standard deviation $\sigma_{\rm OC}$. $RV_{\rm OC}$ was estimated as 
the mean value of the radial velocities of the member stars through an iterative method 
that rejects as outliers those stars that were originally classified as OC members, but 
that have velocities outside the $\pm  3 \sigma_{\rm OC}$-clipping. It was assumed that the 
spectral shift corresponds entirely to the motion of the stars, i.e.\ any other possible 
contributor, such as the gravitational redshift or convective shift, was ignored at this stage. 
This approach is likely to produce a small bias (of a few hundred m\,s$^{-1}$) 
in the estimates of $RV_{\rm OC}$ and $\sigma_{\rm OC}$. The relevance of this minor effect is 
considered in the results and conclusions sections of this paper.

After estimating the radial velocity for each cluster, two additional restriction 
were applied. First, we consider only those clusters with a minimum number of stars 
in order to gain confidence in our estimation of the kinematic parameters of the cluster. 
Several thresholds were considered and $N_{\rm min} \geq 20$ was chosen as a good option to 
keep a relatively large sample. To ensure a tight gravitational binding, those clusters
 with  $\sigma_{\rm OC}> 3$ km\,s$^{-1}$ were also discarded. With these restrictions in place, we were left with $\sim$42\% of the original dataset.

Additional properties of the clusters such as morphology, age, position within our galaxy, 
or the stellar type population, etc., could be also considered as efficient criteria to optimize the sample. However, we decided 
to ignore them to minimize the number of assumptions and to give a more general validity 
to the results. Neither has it been considered the information of tangential velocities 
that should provide a better estimation of the kinematic and dynamical state of the OC. 
That might constitute an additional criteria to determine the membership probability of the 
stars\footnote{A further analysis is possible by comparing tangential and radial velocities of 
the stars, which might provide an alternative way to estimate gravitational redshift.}.

Finally, a total sample of 86 OCs and 8899 stars was selected for further analysis (Table \ref{tab:basesample}). A 
subsample of 51 OCs and 4464 stars are obtained after applying the restrictions based on 
the hydrodynamical model to estimate the convective shift effect of the stars (see below). 
Figure \ref{fig:histRV} depicts the kinematic and descriptive information for both the GR (step 
histogram) and GR+CS (filled histogram) data samples.   
It is expected that the velocity distribution for each cluster follow a Gaussian form, centered 
on $RV_{\rm OC}$ with a dispersion of $\sigma_{\rm OC}$. In order to consider all the clusters, the relative 
radial velocity $\Delta RV$ distribution was depicted in Figure \ref{fig:histRV} (\textit{left 
panel}), subtracting the corresponding $RV_{\rm OC}$. After combining all the clusters, the relative 
velocity histogram fits a single Gaussian distribution centered at 0 with 
$\sigma_{\rm GR} = 1.6$ km \, s$^{-1}$ and $\sigma_{\rm GR+CS} = 1.7$ km \, s$^{-1}$, respectively for GR and GR+CS data. 
The velocity dispersion ranges from $-$8.41 to 8.08 km\,s$^{-1}$ that roughly corresponds to 
the $\pm 3$-sigma level of the dispersion.
The \textit{right panel} shows the histogram distribution of member stars $N$, illustrating
 how populated are the clusters ($N_{\rm OC}$). It serves to compare quantitatively the numbers 
of both data samples.

\subsection{Gravitational Redshift} \label{subsec: GR}

\begin{figure}[tb!]
\makebox[\textwidth][c]{\includegraphics[width=1.05\textwidth]{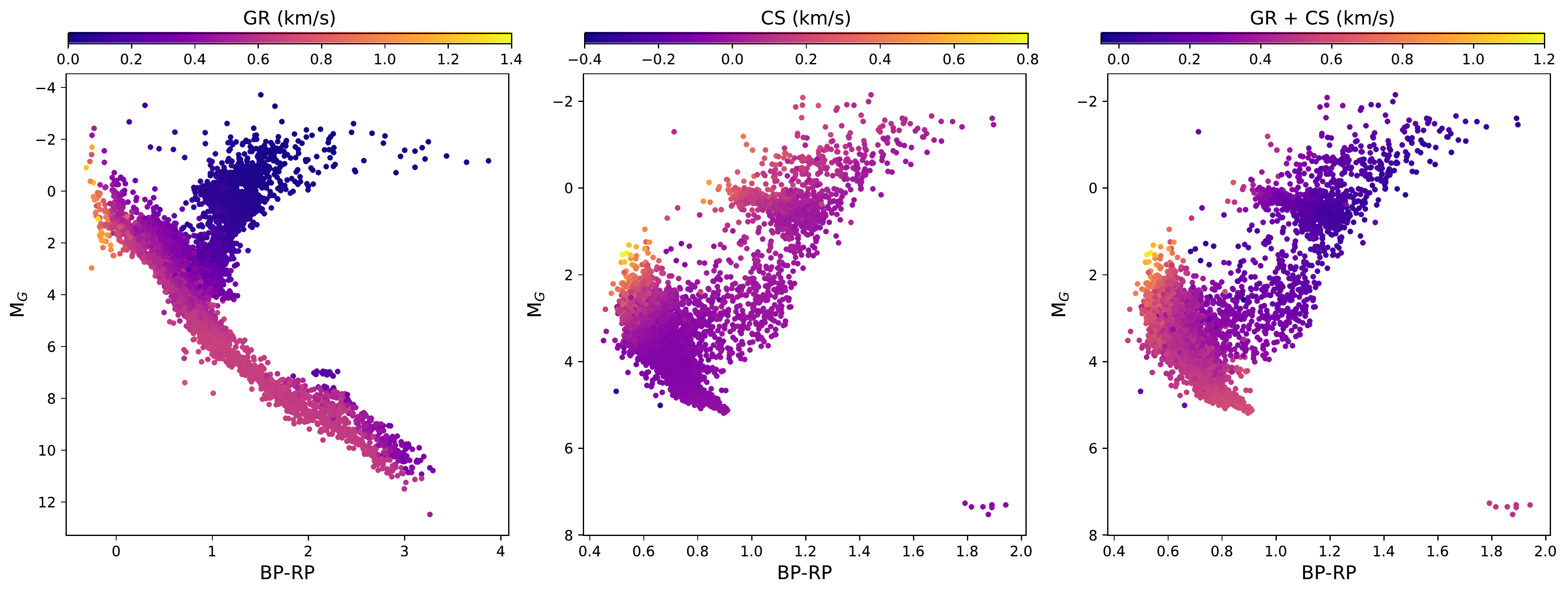}}
\caption{Comprehensive Hertzprung--Russell (HR) diagram with dereddened colors and magnitudes. 
\textit{Left panel}: 86 clusters and 8899 stars, color-coded by the stars' predicted  
gravitational redshift (GR) results (km\,s$^{-1}$). \textit{Middle panel}: 51 clusters 
and 4464 stars, color-coded by the stars' convective shift (CS) results (km\,s$^{-1}$). 
\textit{Right panel}: these 51 clusters and 4464 stars, color-coded by the sum of the
GR and CS results (km\,s$^{-1}$). \label{fig:CMD}}
\end{figure}

To calculate a star's predicted  gravitational redshift (GR) it is necessary to know its mass and 
radius, as $GR \sim M/R$. As a reference value, we estimate the gravitational redshift in the vicinity of the  Sun
 to be ${GR}_\odot = GM_\odot/cR_\odot = 636.31$\footnote{\label{ref_units} Calculated 
using \href{https://docs.astropy.org/en/stable/constants/}{AstroPy} astrophysical constants.} m\,s$^{-1}$.
Therefore, for a given star of mass $M_*$ and radius $R_*$, the predicted gravitational effect is 
given by $GR_*=G(M_*/M_\odot)/c (R_*/R_\odot)$.
The stellar masses are provided by StarHorse catalogue \citep{2019A&A...628A..94A} and their radii 
can be easily estimated through their surface gravity ($\log g \sim M/R^2$), $\log g$ 
also being available for each star. The reference value for $\log g_\odot$ is $4.438^\mathrm{\ref{ref_units}}$.
Figure \ref{fig:CMD}, \textit{left panel}, illustrates the Hertzprung--Russell (HR) diagram for our total 
data sample, color-coded with the GR predictions (km\,s$^{-1}$). The GR values  span a range from a 
minimum of 4.45 m\,s$^{-1}$ for red-giant stars, gradually increasing towards the main-sequence stars
 with a typical value of 0.60 km\,s$^{-1}$ and reaching a maximum of 1.30 km\,s$^{-1}$ for stars above 
the main-sequence turn-off, some of these being known as blue stragglers \citep{2019A&A...627A.119C}.

\subsection{Convective Shift} \label{subsec: CS}
Convective shift (CS) is a hydrodynamic effect in the observed $RV$ due to the convective 
motions of the stellar photosphere \citep{2003A&A...401.1185L}. It produces a net kinematic 
displacement that is reflected in the line-of-sight velocity shift measured for each star.
To estimate the effective contribution of the CS we used the 3D hydrodynamical model developed 
by \citet{2013A&A...550A.103A}, which is described as a function of three stellar parameters; 
namely, the stellar effective temperature ($T_{\rm eff}$), surface gravity ($\log g$), and metallicity 
([Fe/H]). The model is constrained mainly to the main-sequence (MS) and red-giant-branch (RGB) 
stellar parameters. The allowed parameter ranges constitute the additional restrictions on the 
final data set, as mentioned before. The ranges for each parameter are: $T_{\rm eff}[K] \in 
[3790.0, 6730.0]$; $\log g [dex] \in [9.184 \cdot 10^{-4}*T_{\rm eff}-2.482, 4.5]$; [Fe/H][dex]  $ \in [-3.0, 0.0]$. 

After the restrictions were imposed, we were left with a total of 51 clusters and 4464 member stars 
(see Figure \ref{fig:histRV} for comparison with the initial data set). The estimated CS 
contribution per star is color-coded in the \textit{middle panel} of the HR diagram (Figure \ref{fig:CMD}). 
The CS results span a range from a minimum of $-$322.7 m\,s$^{-1}$ (i.e. blueshifted) for a typical sun-like 
star, to a maximum of 804.9 m\,s$^{-1}$ (i.e. redshifted) for the early F-type dwarfs. It is worth
 highlighting that less than 2\% of the stars have a CS greater than 0.3 km\,s$^{-1}$, in agreement 
with the predictions.

The total \emph{velocity shift} contribution to the observed $RV$, estimated from the sum of the GR and CS 
effects, is illustrated in the \textit{right panel} of Figure \ref{fig:CMD}. The results are clearly 
differentiated between the red-giant branch and the main-sequence stars, with a minimum blueshift of 
$-$50.84 m\,s$^{-1}$ and a maximum redshift (for early F-type dwarfs) of 1.16 km\,s$^{-1}$, respectively. 
The majority of the sample, namely 74\% of stars, have a total predicted shift within the range of 0.2--0.7 km\,s$^{-1}$. 

\section{Analysis and Results} \label{sec:analysis}

The basic hypothesis taken into account 
is that the observed spectral shift, in a given star, is a combination of (i) the average radial 
velocity of its OC; (ii) the radial component of the peculiar velocity of the star within its OC; 
(iii) the gravitational redshift ($z_{\rm GR}$) which is equivalent from an observational 
point of view to a radial velocity $GR = c z_{\rm GR}$, and (iv) the effect of the convective shifts (CS) 
due to the stellar turbulence at the photosphere. The large size and the broad range of stellar 
parameters in our sample allow us to consider other second-order effects (e.g., the kinematics of the outer 
layer, orbital motion in binary systems, etc.) as sources  of random noise. The radial velocity of each
 OC was estimated as the average of the radial velocities of it member stars, while the peculiar velocity 
of each star within its own OC was also considered a source of random noise.
In summary, the observed redshift (after subtracting the mean velocity of the OC) is $\Delta RV = GR + 
CS + \sigma $. To test this hypothesis, the data was parameterized as $\Delta RV = a \cdot (GR + CS) + b$, 
where $a$ is the slope correlating the expected results with the observations. Values $a \sim 1$ indicate 
agreement between theory and observations. 

Two different approaches were taken to archive our goal. The first one is based on analyzing a 
fictitious cluster that has been built through the combination of the stellar members of all the OCs, and 
analyze it as a single system (hereafter Method A). The second approach considers only the 
\textit{best}-clusters to analyze them independently (hereafter Method B). In Method A, after subtracting 
the radial velocity of each cluster, we jointly analyze all the stars of host selected clusters with at 
least 20 members, i.e. those stars in the step-histograms shown in Figure 1. Method B analyzes 
OCs with at least 50 members independently. The individual outcome parameters $a$ and $b$ are 
later combined into a \textit{final} result. The analysis and results for each method are described below.
\subsection{Method A}
\begin{figure}[t!]
\plotone{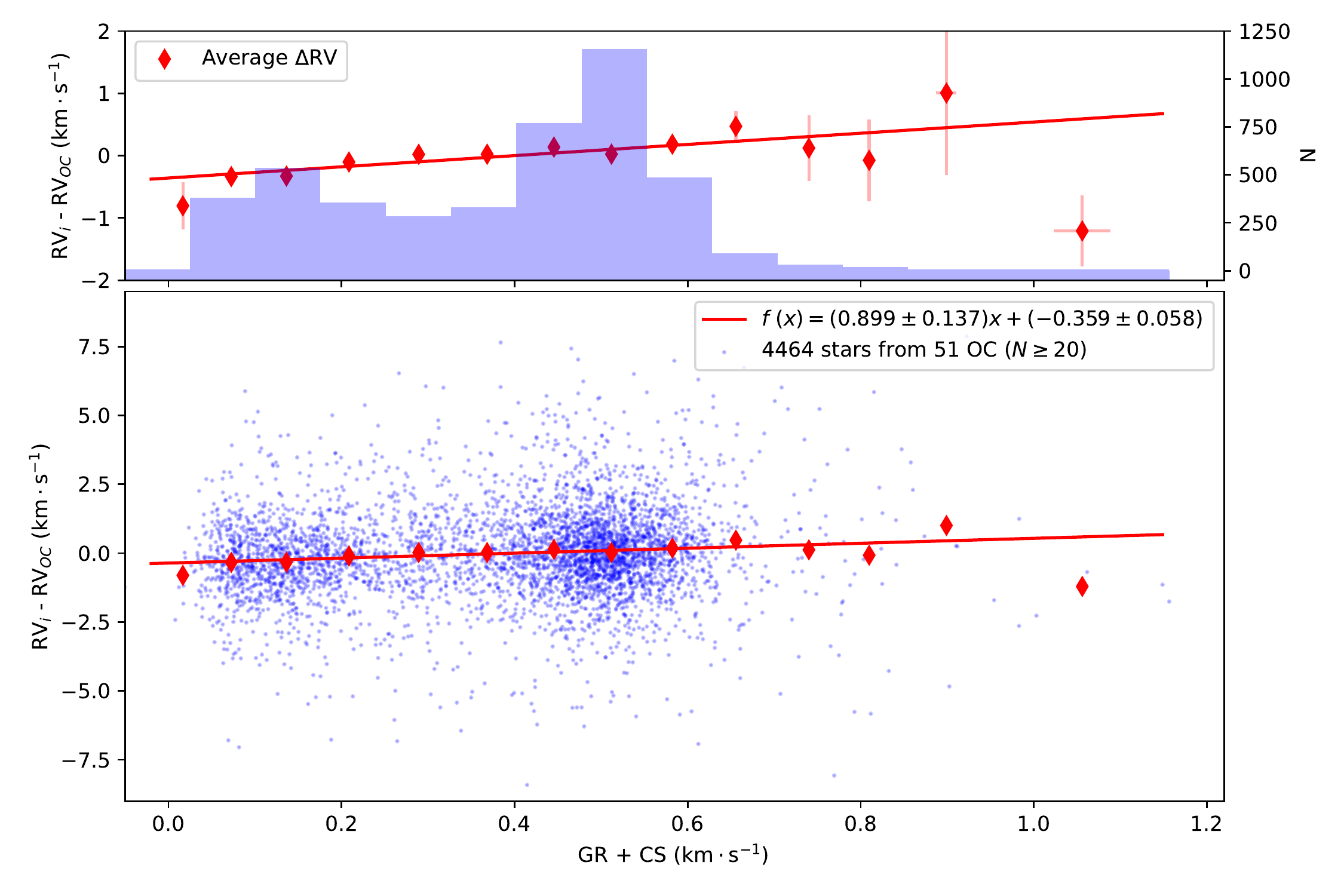}
\caption{\textit{Upper panel}: Bimodal Histogram of GR + CS (km\,s$^{-1}$) for stellar members of 51 OCs 
($N  \geq 20$), where $N$ is the number of the stars (right Y-axis). The 2-peak distribution shows the
 presence of the (two) main stellar groups, i.e. red-giants (left) and main-sequence stars (right). 
\textit{Lower panel}: Linear Regression for $f(x) = \Delta RV$ as a function of $x = $ GR + CS, both in 
(km\,s$^{-1}$). The blue dots represent stellar members of the 51 OCs. In both figures, the red line is
 the result of a least-squares fit to the 4464 stars; the red diamonds are the binned GR + CS data 
corresponding to the average value of $\Delta RV$, shown for illustrative purpose only.\label{fig:OLS1}}
\end{figure}
First, we ignore the CS contribution and consider our primary data set of 86 OCs and 8899 stars.
The aim is to prove, even if without the CS effect, that there is a not-null result when correlating
 $\Delta RV$ with the theoretical GR. Moreover, the CS algorithm was developed specifically for the 
\textit{Gaia} Radial Velocity Spectrometer \citep{2013A&A...550A.103A}, while the $RV$ data used  have 
multiple sources (see ref.\ in \citealt{2021A&A...647A..19T}). To avoid any possible bias from the 
CS algorithm, it is important first  to analyze the GR contribution alone. 
By least-squares fitting, we perform a linear regression for the $\Delta RV$ as a function of GR 
predictions. The results obtained are: $a = 0.425 \pm 0.077$ (slope) and $b = - 0.187 \pm 0.038$ 
(intercept, or the bias term). This results may be interpreted as a significant correlation between 
the observed redshift and the $M/R$ ratio of the stars, although the amplitude of this correlation is 
about half the value predicted by GRT. As in previous studies, we can associate this difference to 
the neglected effect of convective shift. The value of $b$ is interpreted as the mean value of the bias 
associated with the estimation of the radial velocity of each cluster. 

In the following, we include in the analysis the CS contribution of each star using the hydrodynamic 
model proposed by \citeauthor{2013A&A...550A.103A}, as  explained in the previous section. 
Figure \ref{fig:OLS1} (\textit{upper panel}) shows the  histogram of the GR + CS distribution for these 
stars. It exhibits a bimodal shape where the peaks correspond to red-giants (left) and main-sequence stars 
(right). A linear regression was performed for the $\Delta RV$ as a function of the sum of GR and CS 
predictions. Figure \ref{fig:OLS1} (\textit{lower panel}) shows the member stars combined into a single, 
fictitious cluster and the regression line with a slope $a = 0.899 \pm 0.137$, and intercept $b = -0.359
 \pm 0.058$. The value obtained for $a$ is compatible with unity and is therefore consistent with the 
joint predictions of the GTR and convective shift model. The value of the $b$ parameter may be interpreted
 as the shift due to the average bias in the estimates of the radial velocities of the clusters 
(calculated ignoring the GR + CS term of each star). In fact, the mean $\overline{GR + CS}$ for the whole
 data set (4,464 objects) is $0.383$ km\,s$^{-1}$, compatible with the absolute value of $|b| = 0.359$ km\,s$^{-1}$. 

One of the biggest limitations of this model is the systematic bias that could arise when different 
types of clusters are combined; for example, combining different clusters having only one-type 
stellar populations (i.e., either red giants or main-sequence stars). In that case, when subtracting the  
$RV_{\rm  OC}$ (defined as the mean of $RV_{i}$, for all \textit{i}-member stars) and afterwards combining 
the clusters, any substantial difference between the two stellar types is ruled out. Increasing the 
number of stars per cluster ($N$) should avoid this systematic bias. Actually, performing a quick analysis
 varying the number of stars per combined cluster, we realized that the slope $a$ converges to unity with 
increasing $N$. A more conservative analysis would  be to perform a linear fit per cluster with a higher $N$
 threshold, and then combine the slope results. This analysis has been done in the following section.   

\subsection{Method B}

The systematic bias introduced in the estimation of the radial velocity of each OC is a major drawback of Method A.
This has motivated the following analysis in which each OC is analyzed individually by estimating the 
parameters $a_j$ and $ b_j$ for each \textit{j}-OC and then combining their results. In this analysis, 
the bias in the estimation of the radial velocity of each OC is absorbed in the parameter $b_j$ and does not affect the slopes $a_j$.

\begin{figure}[t!]
\makebox[\textwidth][c]{\includegraphics[width=\textwidth]{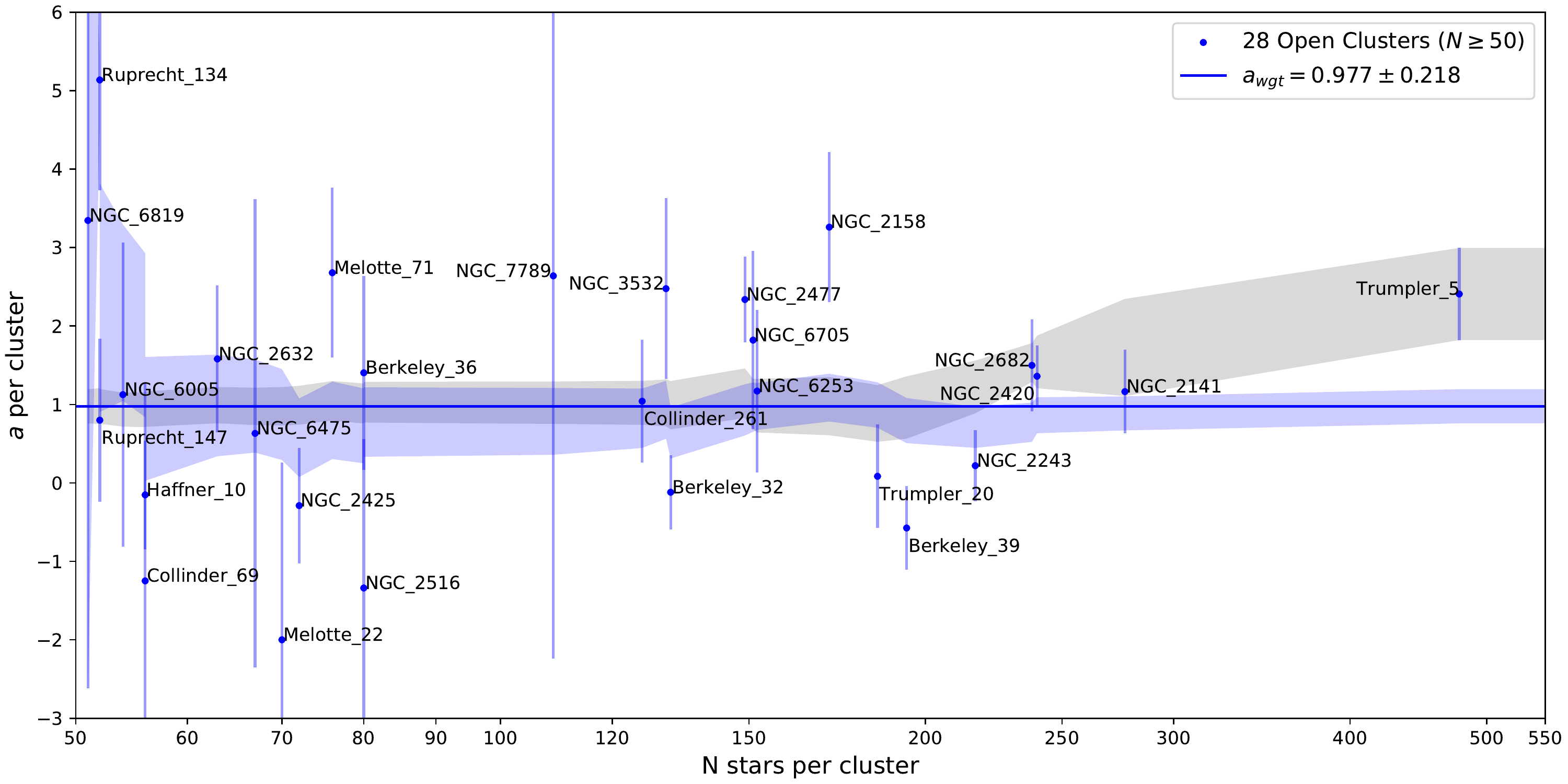}}
\caption{Linear-fit slope ($a$) 
as a function of $N$ stars per cluster. The dashed line 
is the weighted mean slope ($a$) from the 28 OC ($N\geq 50$) results, with a value of 
$0.977 \pm 0.218$. The blue shadow region is the $ \pm \sigma $ cumulative weighted average of each OC 
slope result. Note how the the results converge smoothly to the expected value ($a \sim 1$) with 
increasing $N$. The gray shadow region is the the $ \pm \sigma $ cumulative weighted average with 
decreasing $N$. The error bars are standard deviation errors per cluster. Note that the X-axis is in 
log-scale for visual clarity. \label{fig:OLS2}} 
\end{figure}

Numerous tests were carried out, changing the threshold for the minimum number of stars, particularly 
the $N\geq 20$ assumed in Method A. The results obtained are robust against the specific  
threshold adopted, although the parameters obtained for OCs with relatively low numbers of stars are
 very noisy and contribute little to the average value. Therefore, a higher limit was applied for
 this analysis. The following results correspond to a subsample of OCs with at least 50 stars
 ($N \geq 50$). This restriction reduced the original GR-sample to the \textit{best} 47 OCs with 7713 stars.
When the CS contribution is included in the analysis, the complete GR + CS sample is reduced to 28 OCs with 3779 member stars.  

The $a_j$ and $b_j$ parameters obtained in the analysis of each OC were weighted according to their 
respective errors $\varepsilon_{a_j}$ and $\varepsilon_{b_j}$. The averaged results, ignoring the CS 
contribution, are $a = 0.643 \pm 0.164$ and $b = -0.257 \pm 0.067$. The $a$ value obtained is consistent
 with the non-null hypothesis at $\sim$$4 \sigma$ level. Modeling the CS effect, the weighted average 
results are $a = 0.977 \pm  0.218$ and $b = -0.395 \pm 0.082$.

Figure \ref{fig:OLS2} and Table \ref{tab:results}
 show the $a_j$ results and their standard errors for each OC when considering the CS contribution (the \textit{best}-cluster subsample). 
The blue- and gray-shaded regions enclose the $a \pm \varepsilon_a$ values obtained from the 
cumulative mean along the increasing (blue) and decreasing (gray) $N$ values. 
In both cases it is worth noting the progressive and smooth convergence to the predicted value $a=1$. 
Moreover, splitting the data between the most and least populated clusters, should reproduce compatible 
results within the expected outcome. Combining only the first 18 OCs ($N \leq 150$), we get 
$a = 1.14 \pm 0.32$, while the last 10 OCs ($N > 150$) give $a = 1.01 \pm 0.32$. Therefore, the 
analysis of two independent data sets  successfully reproduces the predictions.

Although the results for all the OCs show compatibility within the $\pm 3\sigma$ with the $a=1$ value, 
the uncertainties in their respective estimations are relatively large and render it impossible to
reject the null hypothesis in the analysis of single OCs. 
Interestingly, the result found in the widely analyzed M67 (NGC 2682) cluster ($ a = 1.499 \pm 0.587$, or $1.178 \pm 0.388$ without modelling the CS contribution) is in tension with previous studies \citep{2011A&A...526A.127P}.

In general, the error bars show a strong 
anticorrelation with the number of stellar members of each OC. However, some of the cases show 
comparatively higher than expected slopes with low error bars (i.e., Trumpler 5). This could be a consequence 
of a low ratio of dwarf or giants, kinematic segregation between types, or irregularities in their spatial 
or kinematic distributions. These hypothesis will require further in-depth study that is beyond the scope of this paper.

\begin{deluxetable*}{lCLCLlRLC}[htb!]
\tablecolumns{9}
\tablewidth{0pt}
\tablecaption{Fit results and parameters$^i$ for OCs with at least 50 members. \label{tab:results}}
\tablehead{   
 \colhead{\vspace{-0.2cm} Cluster } &  \colhead{a} & \colhead{$\varepsilon_{a}$} & \colhead{b} & \colhead{$\varepsilon_{b}$} & \colhead{N} & \colhead{$RV_{OC}$} & \colhead{$\varepsilon_{RV_{OC}}$} & \colhead{$\sigma_{OC}$}
 \vspace{-0.2cm}
 \\ 
  } 
\startdata
   Trumpler\_5 &  2.409 &  0.587 & -1.014 &  0.268 & 478 &  51.889 &  0.095 & 2.293 \\
     NGC\_2141 &  1.164 &  0.533 & -0.479 &  0.229 & 277 &  26.802 &  0.078 & 1.462 \\
     NGC\_2420 &  1.360 &  0.393 & -0.671 &  0.187 & 240 &  74.939 &  0.049 & 0.839 \\
     NGC\_2682 &  1.499 &  0.587 & -0.640 &  0.251 & 238 &  34.451 &  0.063 & 1.093 \\
     NGC\_2243 &  0.220 &  0.452 & -0.117 &  0.200 & 217 &  59.929 &  0.064 & 1.048 \\
  Berkeley\_39 & -0.575 &  0.532 &  0.128 &  0.215 & 194 &  59.053 &  0.079 & 1.287 \\
  Trumpler\_20 &  0.084 &  0.660 &  0.043 &  0.284 & 185 & -39.553 &  0.110 & 1.973 \\
     NGC\_2158 &  3.261 &  0.959 & -1.167 &  0.383 & 171 &  28.678 &  0.186 & 2.694 \\
     NGC\_6253 &  1.171 &  1.036 & -0.395 &  0.433 & 152 & -28.375 &  0.102 & 1.549 \\
     NGC\_6705 &  1.820 &  1.138 & -1.083 &  0.536 & 151 &  36.081 &  0.149 & 2.597 \\
     NGC\_2477 &  2.338 &  0.547 & -0.715 &  0.192 & 149 &   8.155 &  0.099 & 1.330 \\
  Berkeley\_32 & -0.119 &  0.472 &  0.017 &  0.192 & 132 & 106.216 &  0.069 & 0.863 \\
     NGC\_3532 &  2.477 &  1.154 & -1.265 &  0.587 & 131 &   5.803 &  0.067 & 1.479 \\
Collinder\_261 &  1.041 &  0.781 & -0.243 &  0.244 & 126 & -24.423 &  0.094 & 1.309 \\
     NGC\_7789 &  2.640 &  4.879 & -0.192 &  0.437 & 109 & -54.134 &  0.124 & 1.499 \\
  Berkeley\_36 &  1.404 &  1.237 & -0.408 &  0.390 &  80 &  63.235 &  0.163 & 1.704 \\
     NGC\_2516 & -1.340 &  1.900 &  1.002 &  0.998 &  80 &  24.200 &  0.054 & 1.067 \\
   Melotte\_71 &  2.680 &  1.083 & -1.133 &  0.482 &  76 &  51.811 &  0.173 & 1.745 \\
     NGC\_2425 & -0.289 &  0.737 &  0.033 &  0.344 &  72 & 104.085 &  0.113 & 1.155 \\
   Melotte\_22 & -2.000 &  2.258 &  1.283 &  1.201 &  70 &   6.077 &  0.051 & 0.998 \\
     NGC\_6475 &  0.632 &  2.986 & -0.516 &  1.483 &  67 & -14.717 &  0.231 & 2.180 \\
     NGC\_2632 &  1.581 &  0.936 & -1.102 &  0.482 &  63 &  35.459 &  0.071 & 0.963 \\
 Collinder\_69 & -1.249 &  2.518 &  0.942 &  0.965 &  56 &  28.388 &  0.170 & 1.744 \\
   Haffner\_10 & -0.153 &  0.695 & -0.002 &  0.285 &  56 &  87.999 &  0.114 & 0.946 \\
     NGC\_6005 &  1.126 &  1.942 & -0.203 &  0.810 &  54 & -24.567 &  0.279 & 2.348 \\
 Ruprecht\_147 &  0.800 &  1.042 & -0.323 &  0.534 &  52 &  42.397 &  0.127 & 1.093 \\
 Ruprecht\_134 &  5.138 &  1.404 & -1.141 &  0.418 &  52 & -39.909 &  0.221 & 1.917 \\
     NGC\_6819 &  3.346 &  5.965 & -0.309 &  0.588 &  51 &   2.811 &  0.213 & 1.592 \\
\enddata
\tablenotetext{i}{Values for $b$, $\varepsilon_b$, $RV$, $\varepsilon_{RV}$ and $\sigma_{OC}$ are in km\,s$^{-1}$.}
\tablecomments{The table shows the results obtained for the 28 OCs with at least 50 member stars. OCs are sorted in descending order with N.
Table 1 is published in its entirety in the electronic 
edition of the {\it Astrophysical Journal}.  A portion is shown here 
for guidance regarding its form and content. The full version includes the results without modelling the CS.}
\end{deluxetable*}

\section{Conclusions}

The aim of this paper has been to estimate the gravitational redshift in non-degenerate stars, i.e., the light 
shift when photons escape the gravitational potential of the star, as predicted by the Einstein's Equivalence Principle. This was done by analyzing astrometric, photometric, and spectroscopic information 
provided mainly by the \textit{Gaia} survey of more than 70\,000 non-degenerate stars previously classified 
as members of galactic open clusters. The study also takes into account information on the
spectral shift of the stars (i.e., radial velocities) provided by the ground-based surveys. Radial velocities 
were assumed to be a combined result of kinematic factors (motion of the OC, peculiar velocity of the stars 
within its own OC, and convective shift) and the gravitational redshift. When considering 
each OC, the membership of the stars was re-estimated on the basis of their kinematic information. To ensure
the general validity of the results, a minimum number of restrictions on the properties of the OCs and their 
members was assumed. That gives a baseline sample that contains 86 clusters and 8899 member stars. The 
main conclusions of the study are the following:

\begin{itemize}
\item Ignoring the effect of convective shift, the estimated redshift, which depends on the M/R ratio, is lower in amplitude by roughly a factor of 1.5 than the theoretical expectations from EEP. Similar results were obtained regardless the applied method (A, B).

\item By estimating the contribution of convective shift, the original sample is reduced to 51  
OCs with more than 20 members each, and a total of 4464 stars. The results show a correlation between 
predictions and observations as quantified by a significant correlation $a = 0.899 \pm 0.137$. 
This value is compatible to within 1$\sigma$  with the predictions (i.e. $a = 1$). 
\item An improved method applied to the \textit{best}-cluster subsample containing 28 OCs ($N \geq 50$ 
each) and 3779 stars gives a correlation in accordance with the predictions; namely, $a = 0.977 \pm 0.218$. 
\item When splitting the complete GR + CS sample of 28 OCs into two independent sets -- (i) OCs with lower
 number of members ($N \leq 150$ per cluster) and (ii) OCs with higher number of members ($N > 150$ per cluster) -- the results are $a = 1.14 \pm 0.32$ and $a = 1.01 \pm 0.32$, respectively. This latter approach gives further confidence in the overall results.

\end{itemize}

In summary, the result is robust since it was extensively tested on a variety of constraints and restrictions
 used to form the sample. This constitutes one of the widest and more conclusive estimations of gravitational redshift in non-degenerate
 stars. The result shows the 
 potential of large astronomical observations in testing physical theories and is a step forward toward proving the universal 
character of physical laws. Potentially, incorporating information on tangential velocities jointly 
with the radial velocities provided by \textit{Gaia}-DR3, and other ground-based surveys, will significantly
 enhance the sample. A better and wider spatial sampling will add further significance to the result of this work.

\begin{acknowledgments}
We thank Y. Tarricq for kindly providing the $RV$ data prior to the public release. We also thank
 M. L\'opez-Corredoira and E. Mediavilla for useful comments and insights into several aspects such as methodology and interpretation  of this study. We would like also to acknowledge an anonymous referee for his/her useful comments, and in particular those clarifying the scope of the results. This work has made use of data from the European Space Agency (ESA) mission
{\it Gaia} (\url{https://www.cosmos.esa.int/gaia}), processed by the {\it Gaia}Data Processing and Analysis Consortium (DPAC,
\url{https://www.cosmos.esa.int/web/gaia/dpac/consortium}). Funding for the DPAC has been provided 
by national institutions, in particular the institutions participating in the {\it Gaia} Multilateral 
Agreement. This work was funded by the Agencia Estatal de Investigación (AEI), of the spanish Ministerio de Ciencia e Innovación (MICINN) under the project PID2019-110614GB-C21/AEI/10.13039/501100011033. N.R. also acknowledges support from the AEI  and the European Social Fund (ESF) under grant \textit{Ayuda para 
contratos predoctorales para la formación de doctores} with reference PRE2020-095880.
\end{acknowledgments}

\facilities{\textit{Gaia} DR2 \citep{2016A&A...595A...1G, 2018A&A...616A...1G}.}

\software{Astropy \citep{astropy:2013, astropy:2018}, Pandas \citep{reback2020pandas}, 
Matplotlib \citep{Hunter:2007}, Statsmodels \citep{seabold2010statsmodels}, AdjustText (\url{https://github.com/Phlya/adjustText}).   
         }
         
\newpage
\appendix 

Table 2 provides the information about the radial velocity and the estimation of the GR and CS effect for 8899 stars in 86 clusters. 

\begin{deluxetable*}{lrrlrrrr}[htb!]
\tablewidth{0pt}
\tablecaption{The base sample. \label{tab:basesample}}
\tablehead{   
\colhead{Gaia DR2} & \colhead{RA} & \colhead{Dec} & \colhead{Cluster} & 
\colhead{RV\tablenotemark{a}} & \colhead{$\sigma_{RV}$\tablenotemark{a}} & 
\colhead{CS} & \colhead{GR} \\
\colhead{} & \colhead{(J2000)} & \colhead{(J2000)} & \colhead{} &
\colhead{(km s$^{-1}$)} & \colhead{(km s$^{-1}$)} & 
\colhead{(km s$^{-1}$)} & \colhead{(km s$^{-1}$)} 
} 
\startdata
  38354680725946240 &58.75658414  & 12.485542172& Melotte\_25 &    36.265&0.172&        &0.6780  \\
  43538293935879680 &58.77775368  & 16.998372278& Melotte\_25 &    36.509&0.064&        &0.6675 \\
  44894472808190720 &55.48191696  & 18.759813169& Melotte\_25 &    35.201&0.695&        &0.6357 \\
  45142206521351552 &61.92550271  & 15.162689108& Melotte\_25 &    38.500  &0.385&  0.0769&0.4145 \\
  45367056650753280 &61.56773091  & 15.698035581& Melotte\_25 &    37.602&0.184& -0.0881&0.6148 \\
\enddata
\tablenotetext{a}{From Tarricq et al. (2021).}
\tablecomments{Table 2 is published in its entirety in the electronic 
edition of the {\it Astrophysical Journal}.  A portion is shown here 
for guidance regarding its form and content.}
\end{deluxetable*}

\bibliography{paper_feb22}
\bibliographystyle{aasjournal}

\end{document}